\documentclass[aps,prb,twocolumn,superscriptaddress]{revtex4-2}
\usepackage{float}
\usepackage{amsmath}
\usepackage{mathtools} 
\usepackage{amssymb}
\usepackage{amsthm}
\usepackage{bm} 
\usepackage{upgreek} 
\usepackage{xcolor}
\usepackage{graphicx}
\usepackage{subfigure}
\usepackage{epstopdf}
\usepackage[colorlinks,linkcolor=blue,anchorcolor=blue,citecolor=blue,urlcolor=blue]{hyperref}
\usepackage{braket}
\usepackage{orcidlink}
\usepackage{setspace}
\usepackage{chemformula}

\begin{document}
\begin{spacing}{1.0}
\title{Two-gap to Single-gap Transition and Two-dome-like Superconductivity in Alkali-Metal Intercalated Bilayer PdTe$_{2}$}
	
\author{Yu-Lin Han\orcidlink{0009-0001-6308-2977}} 
\affiliation{School of Physics and Physical Engineering, Qufu Normal University, Qufu 273165, China}

\author{Shu-Xiang Qiao} 
\affiliation{School of Physics and Physical Engineering, Qufu Normal University, Qufu 273165, China}
	
\author{Kai-Yue Jiang} 
\affiliation{School of Physics and Physical Engineering, Qufu Normal University, Qufu 273165, China}

\author{Jie Zhang} 
\affiliation{School of Physics and Physical Engineering, Qufu Normal University, Qufu 273165, China}

\author{Bao-Tian Wang} 
\affiliation{Institute of High Energy Physics, Chinese Academy of Sciences, Beijing 100049, China}
\affiliation{Spallation Neutron Source Science Center, Dongguan 523803, China}

\author{Ping Zhang} 
\affiliation{School of Physics and Physical Engineering, Qufu Normal University, Qufu 273165, China}
\affiliation{Institute of Applied Physics and Computational Mathematics, Beijing 100088, China}

\author{C. S. Ting} 
\affiliation{Texas Center for Superconductivity and Department of Physics, University of Houston, Houston TX77204, USA}

\author{Hong-Yan Lu}\email[E-mail:]{hylu@qfnu.edu.cn} 
\affiliation{School of Physics and Physical Engineering, Qufu Normal University, Qufu 273165, China}


\begin{abstract}
PdTe$_{2}$ has been synthesized with controllable thickness down to the monolayer limit. Based on first-principles calculations within the fully anisotropic Migdal-Eliashberg framework, this work reveals that alkali-metal intercalation markedly enhances the weak superconductivity of bilayer PdTe$_{2}$, boosting the transition temperature from 1.4 K to 5.0--13.5 K and yielding a two-dome-like evolution of $T_{c}$. Rubidium intercalation induces the highest $T_{c}$ of 13.5 K, which can be further increased to 14.5 K under biaxial tensile strain. The strain-dependent evolution of $T_{c}$ also exhibits a two-dome-like behavior, reflecting the interplay between strain-induced band structure modifications and electron-phonon coupling (EPC). Moreover, a systematic correlation is identified between interlayer interaction and superconducting gap. Lithium intercalation induces a distinct two-gap state, whereas intercalants with larger atomic radii (Na, K, Rb, and Cs) drive the system into a single-gap character. The two-gap to single-gap transition originates from the modulation of interlayer coupling through intercalation-induced interlayer expansion. In addition, pristine and Li/Na-intercalated bilayers exhibit nontrivial band topology, suggesting that layered PdTe$_{2}$ provides a promising platform for realizing the coexistence of superconductivity and nontrivial topology. These results provide detailed anisotropic insights into EPC and offer viable pathways for enhancing $T_{c}$ and achieving diverse properties in layered PdTe$_{2}$ systems.	
\end{abstract}

\maketitle

\section{Introduction}
Transition-metal dichalcogenides (TMDCs) have emerged as a broad platform for investigating diverse quantum phenomena, encompassing superconductivity \cite{From-Mott-state-to-superconductivity-in-1T-TaS2,Enhanced-superconductivity-in-atomically-thin-TaS2}, charge-density wave (CDW) \cite{Characterization-of-collective-ground-states-in-single-layer-NbSe2}, Dirac semimetals \cite{Lorentz-violating-type-II-Dirac-fermions-in-transition-metal-dichalcogenide-PtTe2}, magnetic order \cite{Van-der-Waals-epitaxial-growth-of-air-stable-CrSe2-nanosheets-with-thickness-tunable-magnetic-order}, and topological states \cite{Ubiquitous-formation-of-bulk-Dirac-cones-and-topological-surface-states-from-a-single-orbital-manifold-in-transition-metal-dichalcogenides,Electrically-switchable-Berry-curvature-dipole-in-the-monolayer-topological-insulator-WTe2}.
In particular, the realization of two-dimensional (2D) superconductivity in TMDCs has attracted growing attention, driven by advances in thin-film growth and device-fabrication techniques \cite{Electronics-and-optoelectronics-of-two-dimensional-transition-metal-dichalcogenides,A-subthermionic-tunnel-field-effect-transistor-with-an-atomically-thin-channel}, which have spurred extensive investigations of 2D crystalline superconductors \cite{Enhanced-superconductivity-in-atomically-thin-TaS2,Characterization-of-collective-ground-states-in-single-layer-NbSe2,Ising-pairing-in-superconducting-NbSe2-atomic-layers,Two-dimensional-superconductivity-and-topological-states-in-PdTe2-thin-films,Synthetic-Control-of-Two-Dimensional-NiTe2-Single-Crystals-with-Highly-Uniform-Thickness-Distributions,Type-II-Ising-Superconductivity-and-Anomalous-Metallic-State-in-Macro-Size-Ambient-Stable-Ultrathin-Crystalline-Films}.
Dimensional reduction often gives rise to unexpected quantum phenomena compared to their bulk counterparts. One paradigmatic case is NbSe$_{2}$, where enhanced CDW and suppressed superconductivity are observed upon reducing the thickness from bulk to the monolayer (ML) \cite{Characterization-of-collective-ground-states-in-single-layer-NbSe2,Stronglyenhancedcharge-density-waveorderinmonolayerNbSe2}, accompanied by a transition from two-gap superconductor to a single-gap one \cite{Quasiparticlespectraof2H-NbSe2Two-bandsuperconductivityandtheroleoftunnelingselectivity,UnusualSuppressionoftheSuperconductingEnergyGapandCriticalTemperatureinAtomicallyThinNbSe2}, reflecting the competition between CDW and superconductivity \cite{Electron-phonon-coupling-and-the-coexistence-of-superconductivity-and-charge-density-wave-in-monolayer-NbSe2,Unveiling-Charge-Density-Wave-Superconductivity-and-Their-Competitive-Nature-in-Two-Dimensional-NbSe2}. 
In addition, although superconductivity in bulk NiTe$_{2}$ has not been detected experimentally \cite{Ionic-liquid-gating-induced-self-intercalation-of-transition-metal-chalcogenides}, a two-gap superconducting state has been predicted to emerge in the ML limit and to be suppressed in bilayer structures \cite{Emergent-superconductivity-in-two-dimensional-NiTe2-crystals}. These findings highlight the crucial role of interlayer interaction in governing the electronic and superconducting properties of TMDCs. 

Interlayer interactions in TMDCs can be effectively modulated by intercalating atoms, molecules, functional groups, or large organic cations into the van der Waals gaps between adjacent layers. This strategy provides a versatile route for tailoring the superconducting properties of both 2D and bulk TMDCs \cite{Ionic-liquid-gating-induced-self-intercalation-of-transition-metal-chalcogenides,Emergent-superconductivity-in-two-dimensional-NiTe2-crystals,Gate-Controlled-K-Intercalation-and-Superconductivity-in-Molybdenum-Disulfide,Enhanced-superconductivity-in-bilayer-PtTe2-by-alkali-metal-intercalations,Superconductivity-in-Li-intercalated-1T-SnSe2-driven-by-electric-field-gating,Induced-anisotropic-superconductivity-in-ionic-liquid-cation-intercalated-1T-SnSe2,Multiple-Intercalation-Stages-and-Universal-Tc-Enhancement-through-Polar-Organic-Species-in-Electron-Doped-1T-SnSe2}. For instance, in K-intercalated 2$H$-MoS$_{2}$, increasing the K concentration via ionic-liquid gating triggers a structural phase transition from the semiconducting 2$H$ phase to the metallic 1$T$ phase, accompanied by the emergence of superconductivity with $T_{c}$ reaching $\sim$ 5 K in K$_{0.4}$MoS$_{2}$ \cite{Gate-Controlled-K-Intercalation-and-Superconductivity-in-Molybdenum-Disulfide}. The gating technique has also been employed to achieve self-intercalation in bulk NiTe$_{2}$ and PdTe$_{2}$ \cite{Ionic-liquid-gating-induced-self-intercalation-of-transition-metal-chalcogenides}, enabling the fabrication of high-quality self-intercalated NiTe$_{2}$ and PdTe$_{2}$ single crystals. In particular, the $T_{c}$ of self-intercalated PdTe$_{2}$ is notably boosted from $\sim$ 1.7 K to 4.3 K \cite{Ionic-liquid-gating-induced-self-intercalation-of-transition-metal-chalcogenides,PdTe-a-strongly-coupled-superconductor,PdTe-a-4.5K-type-II-BCS-superconductor,Experimental-Realization-of-Type-II-Dirac-Fermions-in-a-PdTe2-Superconductor,Type-I-superconductivity-in-the-Dirac-semimetal-PdTe2,Nontrivial-Berry-phase-and-type-II-Dirac-transport-in-the-layered-material-PdTe2,Fermiology-and-Superconductivity-of-Topological-Surface-States-in-PdTe2,Electron-phonon-coupling-in-superconducting-1T-PdTe2}. From a theoretical perspective, alkali-metal intercalation has been predicted to effectively enhance the superconductivity of bilayer NiTe$_{2}$ and PtTe$_{2}$, yielding $T_{c}$ values ranging from 4 to 11 K \cite{Emergent-superconductivity-in-two-dimensional-NiTe2-crystals,Enhanced-superconductivity-in-bilayer-PtTe2-by-alkali-metal-intercalations}. In contrast, in SnSe$_{2}$, intercalation with large organic cations produces an even stronger enhancement of superconductivity than alkali metals, underscoring the role of intercalant size and chemistry in modulating interlayer coupling and electron-phonon interactions \cite{Superconductivity-in-Li-intercalated-1T-SnSe2-driven-by-electric-field-gating,Induced-anisotropic-superconductivity-in-ionic-liquid-cation-intercalated-1T-SnSe2,Multiple-Intercalation-Stages-and-Universal-Tc-Enhancement-through-Polar-Organic-Species-in-Electron-Doped-1T-SnSe2}. Collectively, these findings highlight intercalation as a powerful and flexible approach for probing interlayer interaction mechanisms and realizing tunable superconductivity in TMDC systems.

High-quality PdTe$_{2}$ thin films with thicknesses ranging from ML to 20 MLs have been successfully fabricated on SrTiO$_{3}$ (001) substrates via molecular beam epitaxy (MBE) \cite{Two-dimensional-superconductivity-and-topological-states-in-PdTe2-thin-films,Type-II-Ising-Superconductivity-and-Anomalous-Metallic-State-in-Macro-Size-Ambient-Stable-Ultrathin-Crystalline-Films}. In contrast to its structural analogue NiTe$_{2}$ \cite{Synthetic-Control-of-Two-Dimensional-NiTe2-Single-Crystals-with-Highly-Uniform-Thickness-Distributions,Emergent-superconductivity-in-two-dimensional-NiTe2-crystals}, PdTe$_{2}$ exhibits narrow-gap semiconducting behavior in the ML limit, but undergoes a transition to a metallic phase at the bilayer limit, accompanied by pronounced thickness-dependent evolution of the atomic and electronic properties \cite{Two-dimensional-superconductivity-and-topological-states-in-PdTe2-thin-films}. Despite extensive experimental and theoretical efforts, however, the superconducting $T_{c}$ of the PdTe$_{2}$ system remains relatively low across all thickness regimes \cite{Two-dimensional-superconductivity-and-topological-states-in-PdTe2-thin-films,Ionic-liquid-gating-induced-self-intercalation-of-transition-metal-chalcogenides,Electron-phonon-coupling-in-superconducting-1T-PdTe2,The-occurrence-of-superconductivity-in-sulfides-selenides-tellurides-of-Pt-group-metals,Constitution-and-magnetic-and-electrical-properties-of-palladium-tellurides-PdTe-PdTe2,Superconductivity-in-Cu-Intercalated-CdI2-Type-PdTe2,Protonation-enhanced-superconductivity-in-PdTe2}. More critically, most prior theoretical studies have characterized its superconducting properties using isotropic models, predominantly based on the McMillan equation \cite{Two-dimensional-superconductivity-and-topological-states-in-PdTe2-thin-films}, leaving the anisotropic superconducting behavior, particularly the momentum-resolved superconducting gap distribution and its temperature-dependent evolution, largely unexplored. These quantities are crucial for unraveling pairing mechanisms in low-dimensional superconductors with anisotropic Fermi surfaces (FSs) \cite{The-origin-of-the-anomalous-superconducting-properties-of-MgB2,Phononic-self-energy-effects-and-superconductivity-in-CaC6,Anisotropic-Migdal-Eliashberg-theory-using-Wannier-functions}. Such considerations, along with the pursuit of 2D superconductors with higher $T_{c}$ and tunable properties, strongly motivate a systematic investigation of the electronic structure and anisotropic superconducting characteristics of layered PdTe$_{2}$.

In this work, we employ first-principles calculations in conjunction with the Wannier interpolation technique to investigate the electronic structure, EPC, and anisotropic superconducting properties of layered PdTe$_{2}$. The results reveal that the weak superconductivity in bilayer PdTe$_{2}$ (henceforth Pd$_{2}$Te$_{4}$) can be significantly enhanced by alkali-metal (group IA) intercalation. The resulting superconducting $T_{c}$ in intercalated Pd$_{2}$Te$_{4}$ exhibits a two-dome-like variation across different intercalants ($\sim$ 5.0$-$13.5 K), with Rb-intercalated Pd$_{2}$Te$_{4}$ (RbPd$_{2}$Te$_{4}$) yielding the highest $T_{c}$ of $\sim$ 13.5 K. Application of biaxial tensile strain further increases the $T_{c}$ of RbPd$_{2}$Te$_{4}$ to 14.5 K, with the strain-dependent $T_{c}$ also exhibiting an intriguing two-dome-like behavior, reflecting strain-induced modifications of the electronic structure and its interplay with phonons. Beyond the overall enhancement of $T_{c}$, by solving the fully anisotropic Migdal-Eliashberg (ME) equations \cite{Anisotropic-Migdal-Eliashberg-theory-using-Wannier-functions}, this work uncovers a systematic correlation between the superconducting gap and interlayer spacing in intercalated Pd$_{2}$Te$_{4}$. Specifically, Li intercalation induces a distinct two-gap superconducting state, whereas for alkali metals with larger atomic radii (Na, K, Rb, and Cs), the superconducting gap reduces to single character. The two-gap to single-gap transition is mainly driven by the modulation of interlayer coupling through intercalation-induced interlayer expansion. Moreover, nontrivial electronic band topology is identified in pristine as well as Li/Na-intercalated Pd$_{2}$Te$_{4}$, suggesting the potential coexistence of superconductivity and topological features in this system. These insights deepen the understanding of EPC in layered PdTe$_{2}$ from an anisotropic perspective and provide valuable avenues for achieving higher-$T_{c}$ superconductivity and diverse properties in low-dimensional quantum materials.

\begin{figure}
	\centering
	\includegraphics[width=1.0\linewidth]{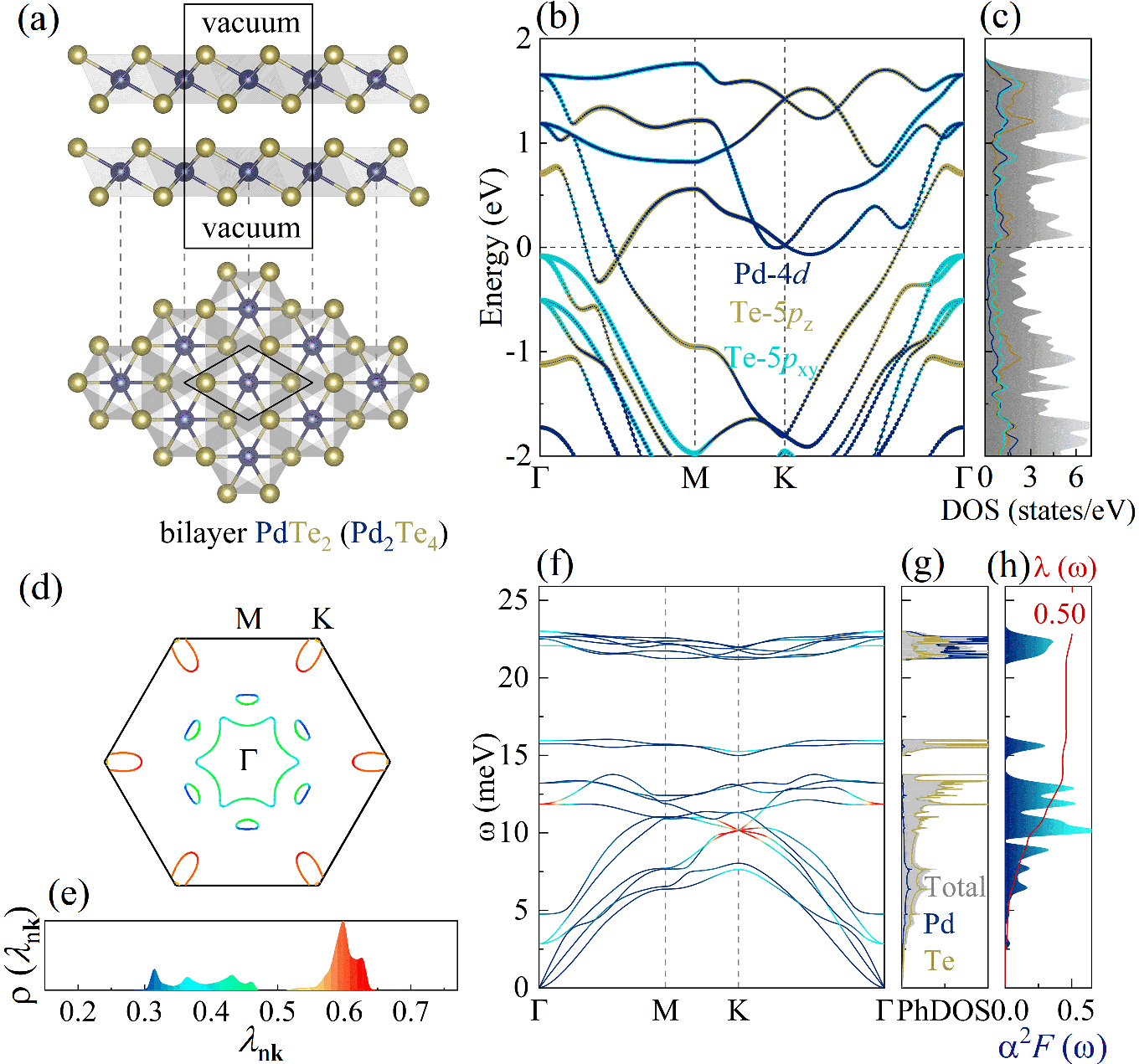}
	\caption{(a) Side and top views of Pd$_{2}$Te$_{4}$. The dark blue and light brown spheres represent Pd and Te atoms, respectively. The 2D hexagonal unit cell is indicated by the black rhombus. (b) Orbital-resolved electronic band structure and (c) density of states (DOS). (d) 2D maps of momentum-resolved EPC strength ($\lambda_{n\mathbf{k}}$) on the FS. (e) Normalized distribution of EPC	strength $\rho$($\lambda_{n\mathbf{k}}$). (f) Phonon dispersion weighted by the $\lambda_{\mathbf{q}\nu}$. (g) Projected phonon DOS (PhDOS). (h) Eliashberg spectral function $\alpha^{2}F(\omega)$ and accumulated EPC constant $\lambda$($\omega$) of Pd$_{2}$Te$_{4}$.}
	\label{fig:fig1}
\end{figure}

\section{Results and discussions}
\subsection{Physical properties of bilayer PdTe$_{2}$}
ML PdTe$_{2}$ consists of a Te-Pd-Te trilayer, in which each Pd is octahedrally coordinated by six Te atoms. The Te atoms occupy the vertices of octahedra, forming a stable 2D sheet with 1$T$-type. Owing to the chemically inert nature of Pd, ML PdTe$_{2}$ can be effectively viewed as a buckled honeycomb structure of Te atoms. Its multilayer and bulk counterparts are formed by AA stacking of such MLs along the $z$ direction \cite{Redetermined-crystal-structures-of-NiTe2-PdTe2-PtS2-PtSe2-and-PtTe2}. The fully optimized lattice constants for ML PdTe$_{2}$ are $a$ = $b$ = 4.03 $\textmd{\AA}$, while those for bulk are $a$ = $b$ = 4.09 $\textmd{\AA}$ and $c$ = 5.19 $\textmd{\AA}$ [Table S1 in Supplemental Material (SM) \cite{Supplemental-Material}], in good agreement with the experimental values ($a$ = $b$ = 4.04 $\textmd{\AA}$ and $c$ = 5.13 $\textmd{\AA}$) \cite{Experimental-Realization-of-Type-II-Dirac-Fermions-in-a-PdTe2-Superconductor,Redetermined-crystal-structures-of-NiTe2-PdTe2-PtS2-PtSe2-and-PtTe2}, given that DFT-PBE typically slightly overestimates lattice constants. In addition, it is noted that the interlayer spacing of bulk PdTe$_{2}$ is about 2.46 $\textmd{\AA}$, which is considerably smaller than that in other layered TMDCs (e.g., 2.89 $\textmd{\AA}$ in WTe$_{2}$ \cite{On-the-Quantum-Spin-Hall-Gap-of-Monolayer-1T-WTe2}, 2.90 $\textmd{\AA}$ in NbSe$_{2}$ \cite{On-the-Structural-Properties-of-the-NbSe2-Phase}, and 3.08 $\textmd{\AA}$ in TiTe$_{2}$) \cite{Emergence-of-charge-density-waves-and-a-pseudogap-in-single-layer-TiTe2}, and even smaller than its structural analogue NiTe$_{2}$ (2.63 $\textmd{\AA}$) \cite{Emergent-superconductivity-in-two-dimensional-NiTe2-crystals}. This suggests potentially strong interlayer coupling in PdTe$_{2}$, which may have profound implications for its electronic structure and superconducting properties, especially in few-layer and intercalated systems.

As mentioned above, ML PdTe$_{2}$ [Fig. S1] \cite{Supplemental-Material} is identified as a narrow-gap semiconductor, consistent with scanning tunneling spectroscopy measurements \cite{Two-dimensional-superconductivity-and-topological-states-in-PdTe2-thin-films}. Given the emergence of metallicity upon layer stacking and the superconducting behavior beyond the ML, our primary focus henceforth shifts to the bilayer form, i.e., Pd$_{2}$Te$_{4}$ [Fig. 1(a)]. The optimized in-plane lattice constants and interlayer spacing for Pd$_{2}$Te$_{4}$ are $a$ = $b$ = 4.06 $\textmd{\AA}$ and $d$ = 2.52 $\textmd{\AA}$, respectively, closely resembling those of the bulk counterpart [Table S1] \cite{Supplemental-Material}. The slight increase in interlayer spacing $d$ from bulk to bilayer reflects a weakening of interlayer coupling. We start by emphasizing the main characteristics of the electronic and phonon band structures of Pd$_{2}$Te$_{4}$. Figures 1(b) and 1(c) display the orbital-resolved band structures and density of states (DOS) (results for ML PdTe$_{2}$ are provided in Fig. S1 \cite{Supplemental-Material}). Upon going from the ML to the bilayer, a pronounced semiconductor-to-metal transition is found, accompanied by strong modification of the electronic structure around the Fermi level ($E_{F}$). The most prominent feature is that a valence band derived mainly from the $p_{z}$-like orbitals of Te atoms is pushed far above the $E_{F}$, giving rise to the metallic character [Fig. 1(b)]. This can be assigned to the strong interlayer hybridization between the $p_{z}$ orbitals of the Te atoms in the bilayer, reflecting the chemically active nature of Te and suggesting strong interlayer coupling. In contrast, the hybridization of $p_{xy}$ bands remains relatively modest. This insight is corroborated by the charge density difference plot in Fig. S3 \cite{Supplemental-Material}, which exhibits significant charge accumulation in the interlayer region. For ML PdTe$_{2}$, by contrast, the crystal-field-induced band splitting between $p_{xy}$ and $p_{z}$ orbitals places $p_{z}$ states below the $p_{xy}$ states, leaving them fully occupied and thereby precluding a metallic state. Additionally, in Pd$_{2}$Te$_{4}$, three bands cross the $E_{F}$ and form multiple Fermi pockets. Centered at $\Gamma$ is a hole pocket derived mainly from the $p$ orbitals of Te, whereas electron pockets at the Brillouin zone (BZ) corners ($K$ points) are dominated by $d$ orbitals of Pd. Another electron pocket lies midway along the $\Gamma$-$M$ direction and exhibits hybridized $d$–$p$ character [Fig. 1(d) and Fig. S7] \cite{Supplemental-Material}. The coexistence of multiple FS sheets with distinct orbital characters and momentum-space anisotropy suggests potentially multi-gap superconducting behaviors in this system.

\begin{figure}
	\centering
	\includegraphics[width=1.0\linewidth]{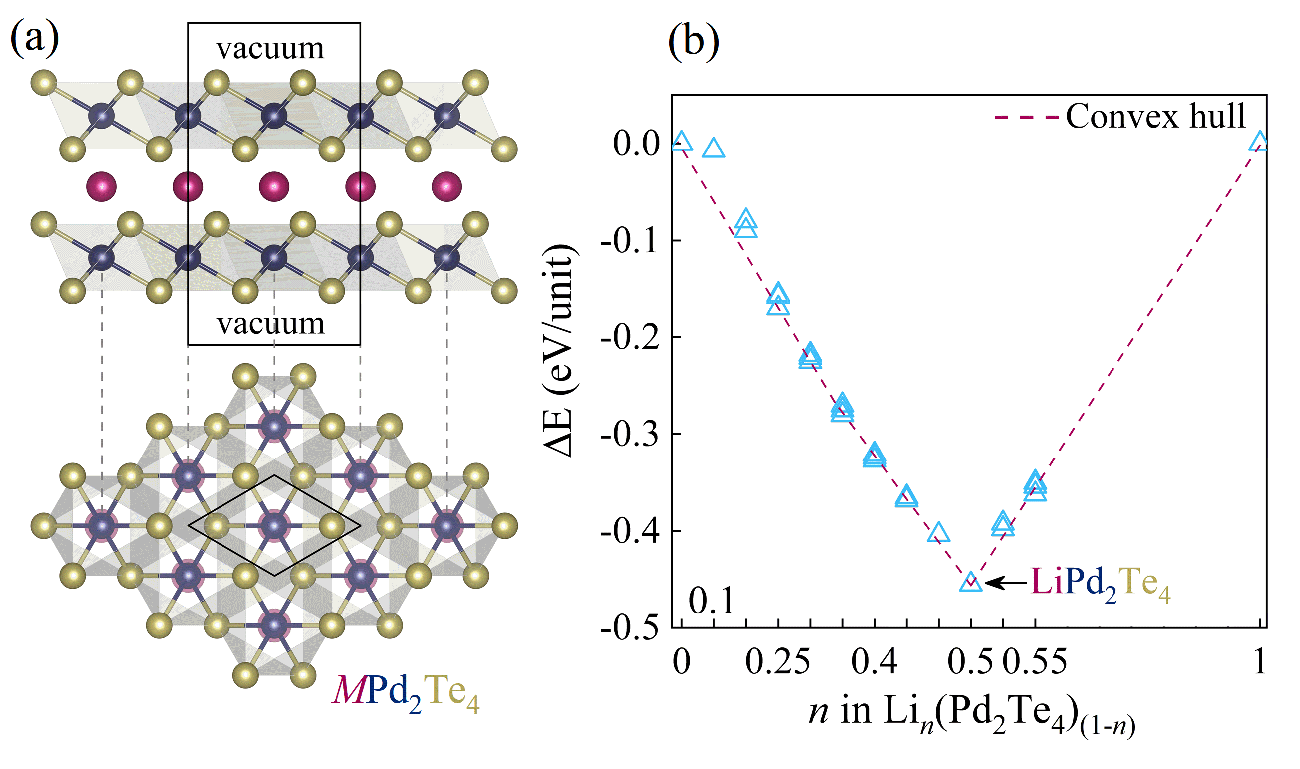}
	\caption{(a) Crystal structure of alkali-metal intercalated Pd$_{2}$Te$_{4}$, in which alkali-metal is represented by the deep pink spheres. (b) Formation energies as a function of alkali-metal concentration ($x$), take Li intercalation for example. The dash line illustrates the convex hull between Pd$_{2}$Te$_{4}$ ($n$ = 0) and the body-centered cubic Li crystal ($n$ = 1).}
	\label{fig:fig2}
\end{figure}

The phonon spectra of ML and bilayer PdTe$_{2}$ [Fig. S1(b) and Fig. 1(f)] \cite{Supplemental-Material} exhibit no imaginary frequencies, confirming their dynamical stability and, in particular, the absence of CDW instabilities that commonly occur in TMDCs \cite{On-the-Structural-Properties-of-the-NbSe2-Phase,Emergence-of-charge-density-waves-and-a-pseudogap-in-single-layer-TiTe2}. The projected phonon density of states (PhDOS) [Fig. 1(g)] exhibits a characteristic separation of vibrational modes into two distinct energy regions separated by a 5 meV gap: low-energy Te-dominated vibrations below 16 meV and high-frequency Pd-dominated modes around 21–23 meV. To gain further insight into the superconducting pairing mechanism, we analyze the phonon dispersion weighted by the mode-resolved EPC strength $\lambda_{\mathbf{q}\nu}$ [Fig. 1(f)] together with the atomic displacement patterns [Fig. S9] \cite{Supplemental-Material}. The results indicate that in-plane vibrations of Te atoms provide the dominant contribution to the EPC. Mapping the momentum-resolved EPC strength $\lambda_{n\mathbf{k}}$ on the FSs reveals pronounced anisotropy, with the strongest coupling concentrated on the $K$-centered electron pocket, whereas the coupling is much weaker on the $\Gamma$-centered hole pocket [Fig. 1(e)]. The discrete distribution of $\rho$($\lambda_{n\mathbf{k}}$) suggests a propensity toward two-gap-like superconductivity. However, the self-consistent solution of the anisotropic ME equations does not converge down to 2 K for Pd$_{2}$Te$_{4}$, indicating that the superconducting $T_{c}$ may lie below this value. Nevertheless, the McMillan-Allen-Dynes (MAD) approach \cite{Transition-Temperature-of-Strong-Coupled-Superconductors,Transition-temperature-of-strong-coupled-superconductors-reanalyzed} yields $T_{c}$ $\sim$ 1.4 K, with EPC constant $\lambda$ = 0.50 and logarithmic average phonon frequency $\omega_{\log}$ = 119.1 K, in good agreement with previous reports \cite{Two-dimensional-superconductivity-and-topological-states-in-PdTe2-thin-films}. As noted above, tuning interlayer coupling has emerged as an effective approach to modulate superconductivity in TMDCs \cite{Ionic-liquid-gating-induced-self-intercalation-of-transition-metal-chalcogenides,Emergent-superconductivity-in-two-dimensional-NiTe2-crystals,Gate-Controlled-K-Intercalation-and-Superconductivity-in-Molybdenum-Disulfide,Enhanced-superconductivity-in-bilayer-PtTe2-by-alkali-metal-intercalations,Superconductivity-in-Li-intercalated-1T-SnSe2-driven-by-electric-field-gating,Induced-anisotropic-superconductivity-in-ionic-liquid-cation-intercalated-1T-SnSe2,Multiple-Intercalation-Stages-and-Universal-Tc-Enhancement-through-Polar-Organic-Species-in-Electron-Doped-1T-SnSe2}. Moreover, our calculations further reveal that electron doping shifts the $E_{F}$ toward the van Hove singularity (vHs) [Fig. 1(c) and Fig. S4], enhancing both the EPC strength and superconducting $T_c$ in Pd$_{2}$Te$_{4}$, whereas hole doping decreases the electronic DOS around the $E_{F}$ and is thus detrimental to superconductivity \cite{Supplemental-Material}. This suggests that alkali-metal intercalation, which simultaneously serves as an electron donor and relieves interlayer coupling by expanding the interlayer spacing, could be a promising route to enhance superconductivity. Motivated by these insights, we investigate the superconducting properties of alkali-metal intercalated Pd$_{2}$Te$_{4}$.

\subsection{Energy favorable structure of intercalated Pd$_{2}$Te$_{4}$}
We first examine the energetically favorable configurations of alkali-metal intercalated Pd$_{2}$Te$_{4}$, taking Li intercalation as an example. The stoichiometry can be expressed as Li$_{x}$(Pd$_{2}$Te$_{4}$)$_{y}$, where $x$ and $y$ are integers denoting the numbers of Li atoms and Pd$_{2}$Te$_{4}$ units, respectively. We adopt a 3 $\times$ 3 $\times$ 1 supercell and evaluate the formation energy $\Delta$$E$ as a function of the Li concentration fraction $n$ = $x$$/$($x$ + $y$), defined as $\Delta$$E(n)$ = $E$[Li$_{n}$(Pd$_{2}$Te$_{4}$)$_{1-n}$] $-$ $n$$E$[Li] $-$ $(1-n)$$E$[Pd$_{2}$Te$_{4}$], where $E$[Pd$_{2}$Te$_{4}$] is the energy of a bilayer PdTe$_{2}$ per Pd$_{2}$Te$_{4}$ unit and $E$[Li] is the energy of a body-centered cubic (bcc) Li crystal per atom.
In experiments, different alkali-metal sources primarily influence the kinetic pathway of intercalation, whereas the thermodynamic stability of the intercalated phase is governed by the chemical potential of the alkali metal, for which bcc Li is adopted as the reference state to calculate the formation energy in this work. 
The corresponding results are presented in Fig. 2(b), where $\Delta$$E(n)$ decreases with increasing $n$ and reaches a minimum at $n$ = 0.5, corresponding to the stoichiometry LiPd$_{2}$Te$_{4}$. Beyond this concentration, the energy increases. The resulting convex hull suggests that LiPd$_{2}$Te$_{4}$ is thermodynamically stable with respect to any other stoichiometry. Accordingly, the same stoichiometry is adopted for the other group-IA alkali-metal intercalations. Figure 2(a) displays the crystal structure of alkali-metal intercalated Pd$_{2}$Te$_{4}$, where the alkali-metal atoms (deep pink spheres) occupy octahedral sites located at the midpoints between nearest-neighbor Pd atoms in adjacent MLs, forming a 2D hexagonal lattice.

\begin{figure}
	\centering
	\includegraphics[width=1.0\linewidth]{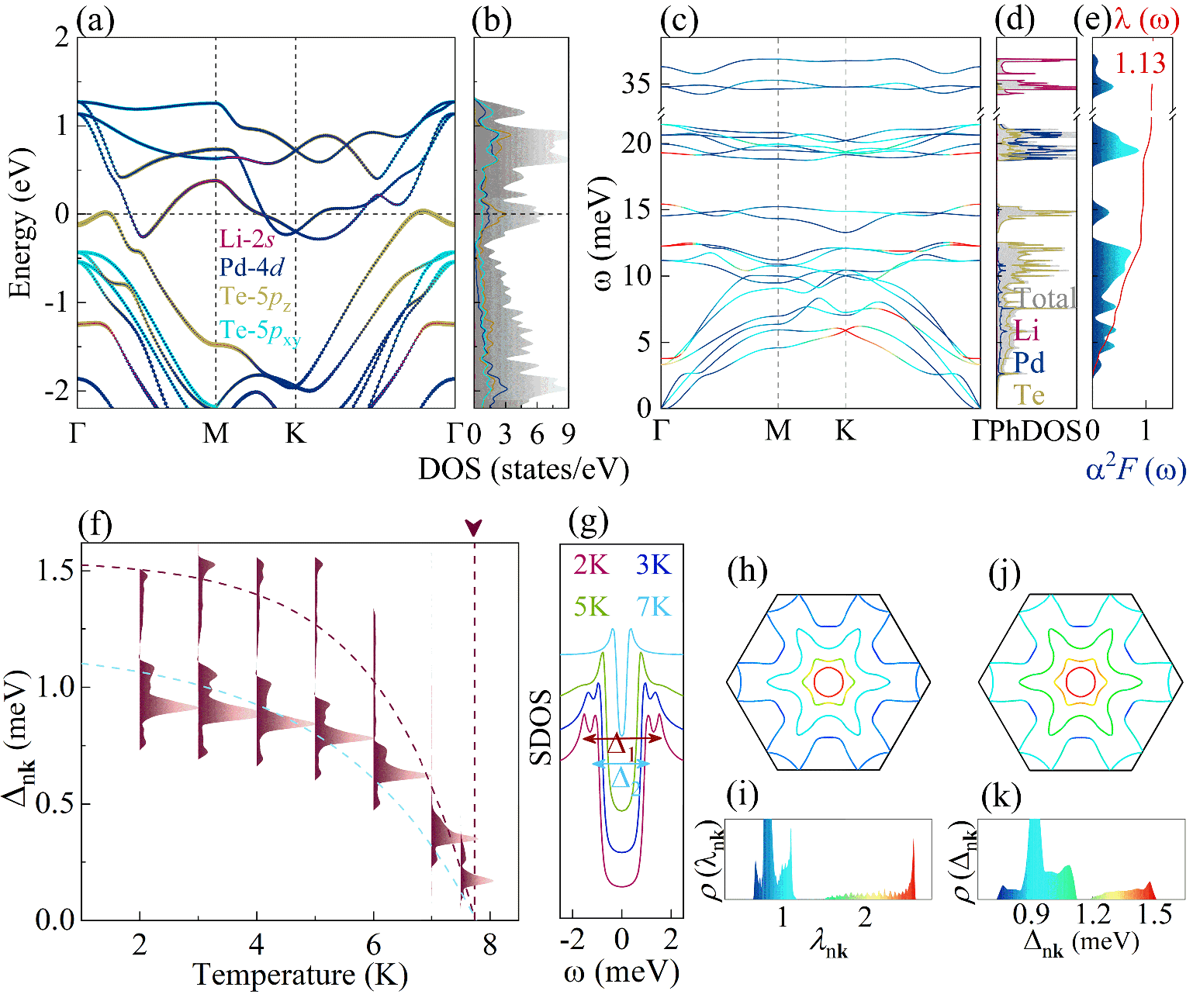}
	\caption{(a)$-$(e) Electronic and phonon properties of LiPd$_{2}$Te$_{4}$. (a) Orbital-resolved band structure. (b) Total and orbital-resolved DOS. (c) Phonon dispersion weighted by the $\lambda_{\mathbf{q}\nu}$. (d) PhDOS. (e) Eliashberg spectral function $\alpha^{2}F(\omega)$ and accumulated EPC constant $\lambda$($\omega$). (f)$-$(j) Anisotropic superconducting properties of of LiPd$_{2}$Te$_{4}$. (f) Energy distribution of the anisotropic superconducting energy gap $\Delta_{n\mathbf{k}}$ as a function of temperature, and the dashed lines serve as guides to the eye. (g) The SDOS under different temperatures of 2, 3, 5, and 7 K, respectively, with the two gaps marked as $\Delta$$_{1}$ and $\Delta$$_{2}$. 2D maps of momentum-resolved (h) EPC strength $\lambda_{n\mathbf{k}}$ and (j) superconducting gap $\Delta_{n\mathbf{k}}$ for each electronic state $n$$\mathbf{k}$ on the FS of LiPd$_{2}$Te$_{4}$. Normalized density of the (i) EPC $\rho$($\lambda_{n\mathbf{k}}$) and (kA) superconducting gap $\rho$($\Delta_{n\mathbf{k}}$).}
	\label{fig:fig3}
\end{figure}

We also examined other candidate interlayer sites for intercalation. The results indicate that the Pd–Pd midpoint octahedral site is energetically most favorable, yielding the lowest total energy among all considered configurations. Detailed comparisons are provided in the Supplemental Materials (SM) \cite{Supplemental-Material}.
$Ab$ $initio$ molecular-dynamics (AIMD) simulations \cite{Efficient-iterative-schemes-for-ab-initio-total-energy-calculations-using-a-plane-wave-basis-set} further confirm that Pd$_{2}$Te$_{4}$ crystals, both pristine and intercalated with Li, Na, K, Rb, or Cs, remain structurally stable at 300 K during a 10 ps simulation, showing no noticeable lattice distortion [Fig. S6] \cite{Supplemental-Material}.
Moreover, the AIMD calculations mainly assess the thermodynamic stability of the intercalated phases under idealized oxygen- and moisture-free conditions and do not directly address oxidation or hydration in ambient environments. Experimentally, intercalated TMDCs are typically handled under an inert atmosphere and protected by encapsulation or passivation \cite{Tailored-Ising-superconductivity-in-intercalated-bulk-NbSe2}. Notably, PdTe$_{2}$ thin films have been reported to exhibit appreciable robustness under ambient exposure. For example, the superconductivity and associated quantum phenomena in macro-scale, ambient-stable ultrathin PdTe$_{2}$ films remain nearly unchanged for up to 20 months \cite{Type-II-Ising-Superconductivity-and-Anomalous-Metallic-State-in-Macro-Size-Ambient-Stable-Ultrathin-Crystalline-Films}. In addition, enhanced superconductivity in 4-MLs PdTe$_{2}$ films has been achieved via magnesium intercalation, suggesting good chemical stability of layered PdTe$_{2}$ under practical handling conditions \cite{Two-dimensional-superconductivity-and-topological-states-in-PdTe2-thin-films,Tailored-Ising-superconductivity-in-intercalated-bulk-NbSe2,High-quality-PdTe2-thin-films-grown-by-molecular-beam-epitaxy,Chemical-Synthesis-and-Integration-of-Highly-Conductive-PdTe2-with-Low-Dimensional-Semiconductors-for-p-Type-Transistors-with-Low-Contact-Barriers}.

\subsection{Two-gap superconductivity in LiPd$_{2}$Te$_{4}$}
Upon Li intercalation, the interlayer spacing expands to about 3.23 $\textmd{\AA}$, which weakens the interlayer coupling and is accompanied by electron transfer from Li to the adjacent Te layers. The synergistic effects between Li-induced interlayer-coupling modulation and electron doping make the band structure of LiPd$_{2}$Te$_{4}$ resemble that of the electron-doped Pd$_{2}$Te$_{4}$, albeit with slight band splitting. As illustrated in Fig. 3(a), despite substantial electron doping into the $e_{g}$ bands, the $t_{2g}$-$e_{g}$ gap remains preserved, and the $p_{z}$-like valence band is shifted downward. Relative to pristine Pd$_{2}$Te$_{4}$, Li intercalation shifts the $E_{F}$ to a vHs, markedly increasing the electronic DOS from 2.75 states/eV in Pd$_{2}$Te$_{4}$ to 6.78 states/eV in LiPd$_{2}$Te$_{4}$. This vHs mainly originates from the dispersionless band along the $\Gamma$--$K$ path, as shown in Figs. 3(a) and 3(b). The increased DOS may contribute to the increased availability of electronic states conducive to superconducting states. Meanwhile, similar to pristine Pd$_{2}$Te$_{4}$, LiPd$_{2}$Te$_{4}$ also hosts three bands crossing $E_{F}$. Specifically, the first two sheets centered at $\Gamma$ are derived from the $p_{z}$-like valence band, comprising an inner electron pocket and an outer hole pocket. The remaining sheets are formed by conduction bands dominated by $d$ orbitals of Pd [Fig. S7] \cite{Supplemental-Material}.

\begin{figure}
	\centering
	\includegraphics[width=1.01\linewidth]{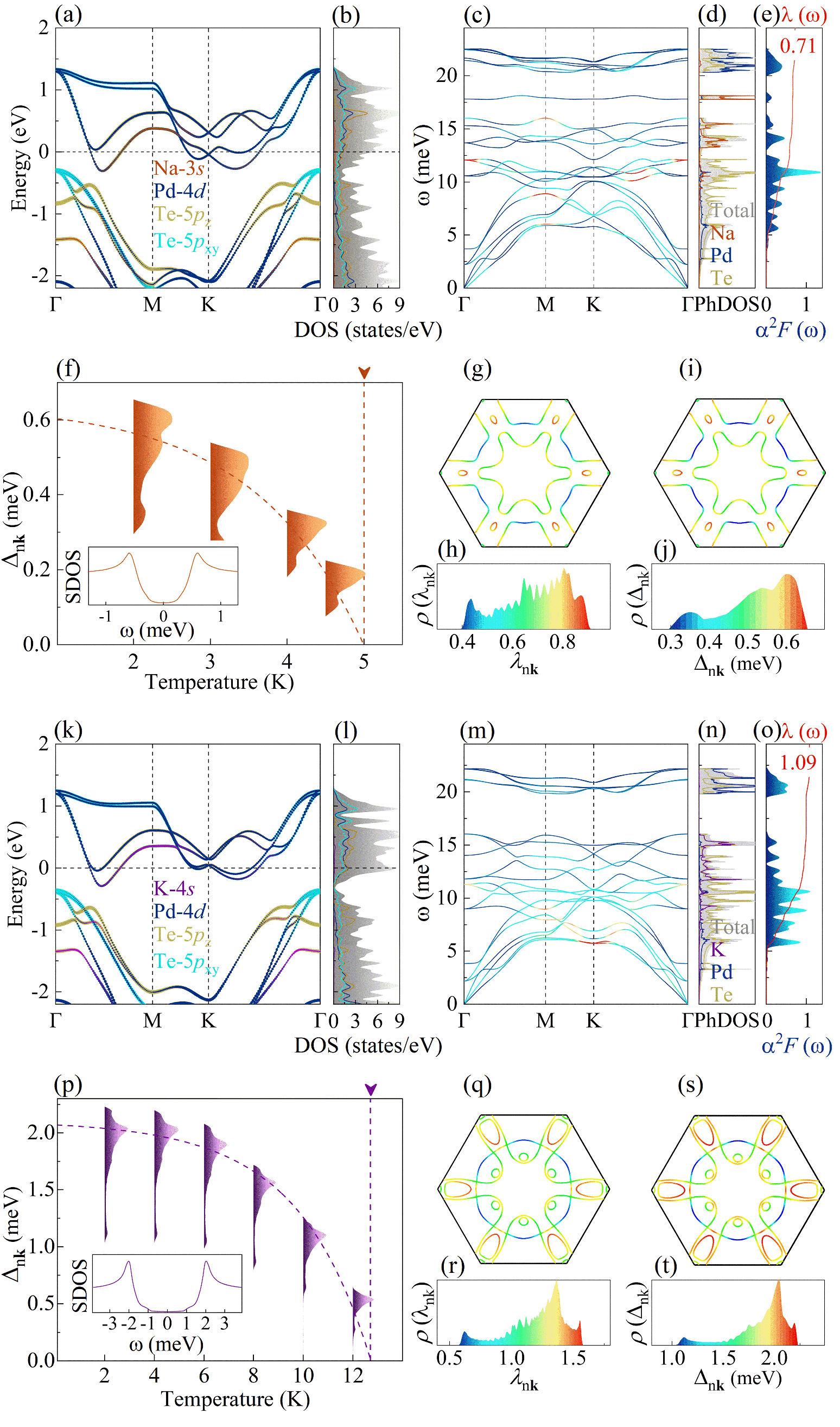}
	\caption{(a)$-$(e) Electronic and phonon properties of NaPd$_{2}$Te$_{4}$. (a), (b) Orbital-resolved band structure and DOS. (c) Phonon dispersion weighted by the $\lambda_{\mathbf{q}\nu}$. (d) PhDOS. (e) Eliashberg spectral function $\alpha^{2}F(\omega)$ and accumulated EPC constant $\lambda$($\omega$). (f)$-$(j) Anisotropic superconducting properties of NaPd$_{2}$Te$_{4}$. (f) Energy distribution of the anisotropic superconducting gap $\Delta_{n\mathbf{k}}$ as a function of temperature, the inset is the SDOS under a low temperature of 2 K. 2D maps of momentum-resolved (g) EPC strength $\lambda_{n\mathbf{k}}$ and (i) superconducting gap $\Delta_{n\mathbf{k}}$ for each electronic state $n$$\mathbf{k}$ on the FS of NaPd$_{2}$Te$_{4}$. Normalized density of the (h) EPC $\rho$($\lambda_{n\mathbf{k}}$) and (j) superconducting gap $\rho$($\Delta_{n\mathbf{k}}$). (k)$-$(t) Corresponding electronic properties and anisotropic superconducting behaviors of KPd$_{2}$Te$_{4}$.}
	\label{fig:fig4}
\end{figure}
In addition to increasing the number of conduction electrons, Li intercalation also significantly enhances the EPC. This is evidenced by the pronounced enhancement of EPC constant $\lambda$ (1.13, with $\omega_{\log}$ of 99.4 K), as shown in Fig. 3(e), which is more than doubled compared to the pristine bilayer ($\lambda$ = 0.50). The peaks below 22 meV in Eliashberg spectral function $\alpha^{2}F(\omega)$ arise mainly from in-plane vibrations of Te atoms, similar to pristine Pd$_{2}$Te$_{4}$. In addition, the presence of Li, by virtue of its light mass and therefore large dynamical scale, gives rise to energetic phonons at higher frequency around 35 meV. Therefore, even within the isotropic MAD \cite{Transition-Temperature-of-Strong-Coupled-Superconductors,Transition-temperature-of-strong-coupled-superconductors-reanalyzed} estimate, an enhanced superconducting $T_{c}$ is expected. Indeed, the MAD approach yields $T_{c}$ $\sim$ 4.5 K [Table S1], which is an inadequate estimate since the spectrum of momentum-resolved EPC strength $\lambda_{n\mathbf{k}}$ on FS is evidently anisotropic [Fig. 3(g)]. We therefore solve the fully anisotropic ME equations to investigate the anisotropy. 
Figure 3(i) shows that the calculated $\lambda_{n\mathbf{k}}$ and $\Delta_{n\mathbf{k}}$ cluster into two separate ranges. These discrete distributions suggest a prominent two-gap characteristic in LiPd$_{2}$Te$_{4}$, which implies that distinct superconducting gaps may open on different FS sheets.
Figure 3(f) shows the energy distribution of the anisotropic superconducting gap $\Delta_{n\mathbf{k}}$ at different temperatures for LiPd$_{2}$Te$_{4}$, while 3(h) and 3(j) present the 2D maps of momentum-resolved $\lambda_{n\mathbf{k}}$ and $\Delta_{n\mathbf{k}}$ on the FSs.
This interpretation is further supported by the presence of two pairs of quasiparticle peaks in the normalized superconducting density of states (SDOS), as shown in Fig. 3(g). The normalized SDOS 
represents the ratio of the superconducting DOS to the normal-state DOS and can be experimentally measured by scanning tunneling microscopy. 
The two normalized superconducting gaps, $\Delta$$_{1}$ and $\Delta$$_{2}$, are approximately 1.5 and 1.1 meV at temperature of 2 K, respectively. As is clear from the 2D map of $\Delta_{n\mathbf{k}}$ [Fig. 3(h)], the larger gap is mainly distributed on the $\Gamma$-centered sheets derived primarily from the $p_{z}$ orbitals of Te atoms, whereas the smaller gap resides on the remaining sheets, e.g., $M$- or $K$-centered pockets dominated by $d$ orbitals of Pd atoms. 
The similarity between the $\lambda_{n\mathbf{k}}$ and $\Delta_{n\mathbf{k}}$ maps indicates that the anisotropic superconducting gap is closely correlated with the momentum-dependent electron–phonon coupling on Fermi-surface sheets.
The temperature at which $\Delta_{n\mathbf{k}}$ vanishes defines the superconducting $T_{c}$, which is determined to be $\sim$ 7.7 K, with the Coulomb pseudopotential $\mu^{*}_{c}$ set to a typical value of 0.14 \cite{Ab-initio-methods-for-superconductivity}.
Additionally, it is worth noting that the discrete distribution of $\lambda_{n\mathbf{k}}$ on FSs and its normalized density $\rho$($\lambda_{n\mathbf{k}}$) remain qualitatively similar before and after Li intercalation, suggesting that pristine Pd$_{2}$Te$_{4}$ may also exhibit an underlying two-gap tendency, although its $T_{c}$ is much lower.

\begin{figure}
	\centering
	\includegraphics[width=1.01\linewidth]{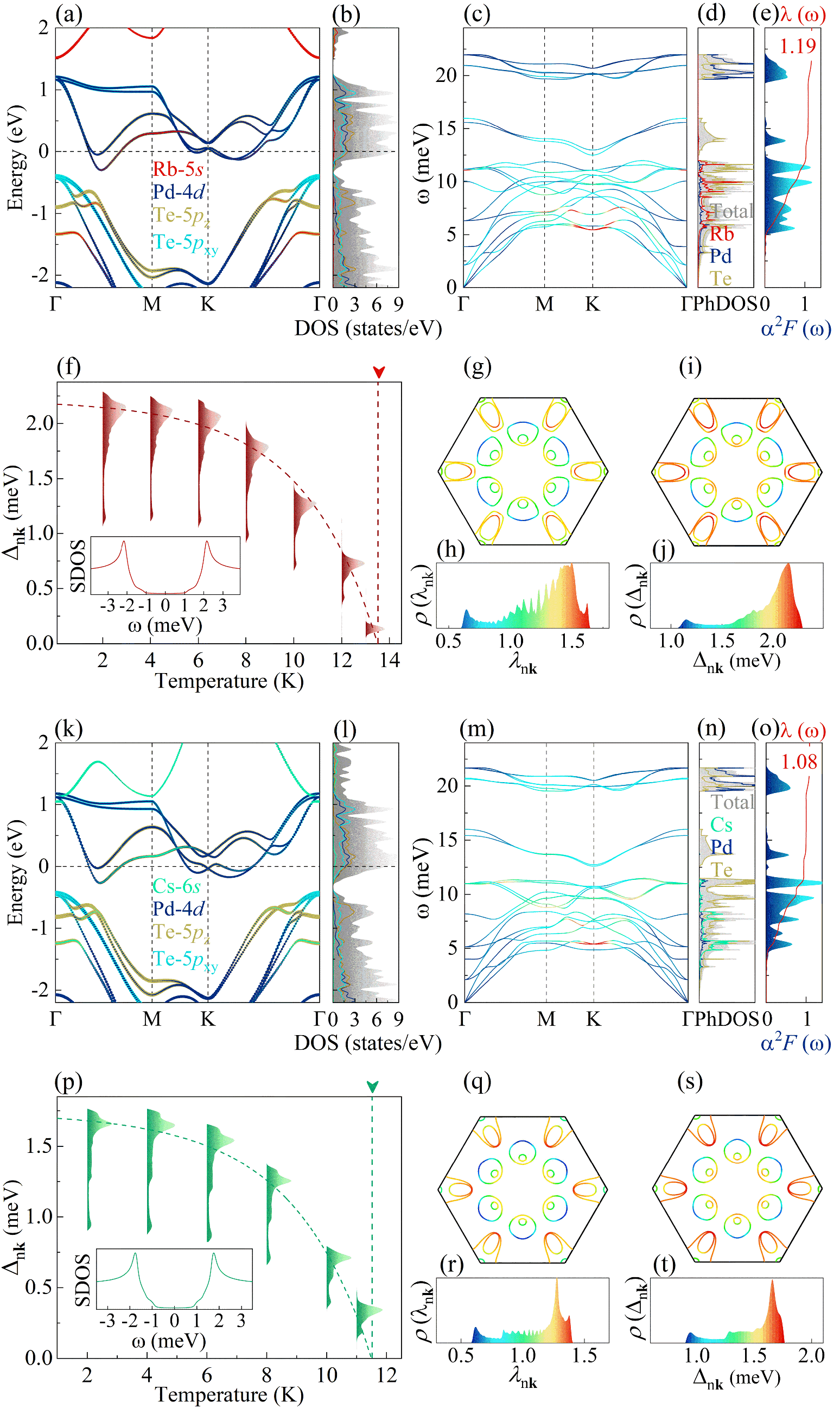}
	\caption{(a)$-$(e) Electronic and phonon properties of RbPd$_{2}$Te$_{4}$. (a), (b) Orbital-resolved band structure and DOS. (c) Phonon dispersion weighted by the $\lambda_{\mathbf{q}\nu}$. (d) PhDOS. (e) Eliashberg spectral function $\alpha^{2}F(\omega)$ and accumulated EPC constant $\lambda$($\omega$). (f)$-$(j) Anisotropic superconducting properties of RbPd$_{2}$Te$_{4}$. (f) Energy distribution of the anisotropic superconducting gap $\Delta$$_{n\mathbf{k}}$ as a function of temperature, the inset is the SDOS under a low temperature of 2 K. 2D maps of momentum-resolved (g) EPC strength $\lambda$$_{n\mathbf{k}}$ and (i) superconducting gap $\Delta$$_{n\mathbf{k}}$ for each electronic state n$\mathbf{k}$ on the FS of RbPd$_{2}$Te$_{4}$. Normalized density of the (h) EPC $\rho$($\lambda$$_{n\mathbf{k}}$) and (j) superconducting gap $\rho$($\Delta$$_{n\mathbf{k}}$). (k)$-$(t) Corresponding electronic properties and anisotropic superconducting behaviors of CsPd$_{2}$Te$_{4}$.}
	\label{fig:fig5}
\end{figure}
\subsection{Single-gap superconductivity in Na, K, Rb, and Cs intercalated Pd$_{2}$Te$_{4}$}	
We next focus on the group-IA intercalants with larger atomic radii (Na, K, Rb, and Cs). Specifically, upon Na intercalation, the interlayer spacing further expands to 4.54 $\textmd{\AA}$. In contrast to pristine and Li-intercalated Pd$_{2}$Te$_{4}$, the $p_{z}$-like valence band that forms the $\Gamma$-centered Fermi pocket no longer crosses the $E_{F}$ in NaPd$_{2}$Te$_{4}$, but shifts downward and becomes fully occupied, such that the metallicity is entirely determined by the conduction bands [Fig. 4(a)]. Meanwhile, the conduction band dominated FSs exhibit stronger $p$-$d$ hybridization than in the pristine and Li-intercalated bilayers. Additionally, the removal of the $p_{z}$-like pocket also leads to a slight reduction of the electronic DOS at $E_{F}$ [Fig. 4(b)] ($N(E_{F})$ = 5.51 states/eV, compared with 2.75 states/eV in Pd$_{2}$Te$_{4}$ and 6.78 states/eV in LiPd$_{2}$Te$_{4}$), which may lead to a decreased $T_{c}$. By integrating $\alpha^{2}F(\omega)$, the obtained EPC constant $\lambda$ and logarithmic average frequency $\omega_{\log}$ of NaPd$_{2}$Te$_{4}$ are $\lambda$ = 0.71 and 110.2 K, as shown in Fig. 4(e). Solving the anisotropic ME equations yields $T_{c}$ $\sim$ 5.0 K, slightly lower than that of LiPd$_{2}$Te$_{4}$,  but still above liquid-helium temperature and much higher than that of pristine Pd$_{2}$Te$_{4}$.

More importantly, the anisotropic results reveal that the absence of the $p_{z}$-like valence band drives a two-gap to single-gap transition in NaPd$_{2}$Te$_{4}$, as presented in Figs. 4(f)--4(j). This transition manifests as a more continuous distribution of $\lambda_{n\mathbf{k}}$ and $\Delta_{n\mathbf{k}}$, in sharp contrast to the discrete distributions in pristine and Li-intercalated cases, and is further corroborated by the evolution of the SDOS from two peaks to a single peak [Fig. 4(f)]. 
This transition can be traced to the sensitivity of the $p_{z}$-like valence band to the interlayer distance between the two adjacent PdTe$_{2}$ MLs. To illustrate this point more clearly, we calculated the electronic band structures of a PdTe$_{2}$ bilayer model while systematically increasing the interlayer spacing from $d$ to $d+3.0~\textmd{\AA}$ in steps of 0.2 $\textmd{\AA}$ ($d$ = 2.52 $\textmd{\AA}$ is the interlayer spacing of pristine Pd$_{2}$Te$_{4}$) [Fig. S11(a)--S11(p)] \cite{Supplemental-Material}. The results indicate that, as the interlayer distance increases, the electronic band structure of Pd$_{2}$Te$_{4}$ progressively approaches that of ML PdTe$_{2}$, and its metallicity is largely suppressed at $d$ + 2.0 $\textmd{\AA}$ [Fig. S11(k)] \cite{Supplemental-Material}. Moreover, the $p_{z}$-like valence band, which contributes to the larger superconducting gap in LiPd$_{2}$Te$_{4}$, no longer crosses the $E_{F}$ once the interlayer spacing is increased by $\sim$ 1.2 $\textmd{\AA}$ [Fig. S11(g)] \cite{Supplemental-Material}. Since Li intercalation increases the interlayer spacing by only $\sim$ 0.7 $\textmd{\AA}$, which is below this threshold (1.2 $\textmd{\AA}$), the two-gap character is preserved. By contrast, Na intercalation induces a much larger expansion of the interlayer spacing ($\sim$ 2.0 $\textmd{\AA}$), thereby driving the two-gap to single-gap transition. 
In addition, a direct comparison between the electronic structures of pristine Pd$_{2}$Te$_{4}$ at different interlayer distances and those of the alkali-metal intercalated systems reveals a key distinction: for the same degree of interlayer expansion, the alkali-metal intercalated Pd$_{2}$Te$_{4}$ invariably remains metallicity. This highlights that the role of alkali metals is twofold, acting as interlayer spacing regulators while simultaneously serving as electron donors.
\begin{figure}
	\centering
	\includegraphics[width=1.0\linewidth]{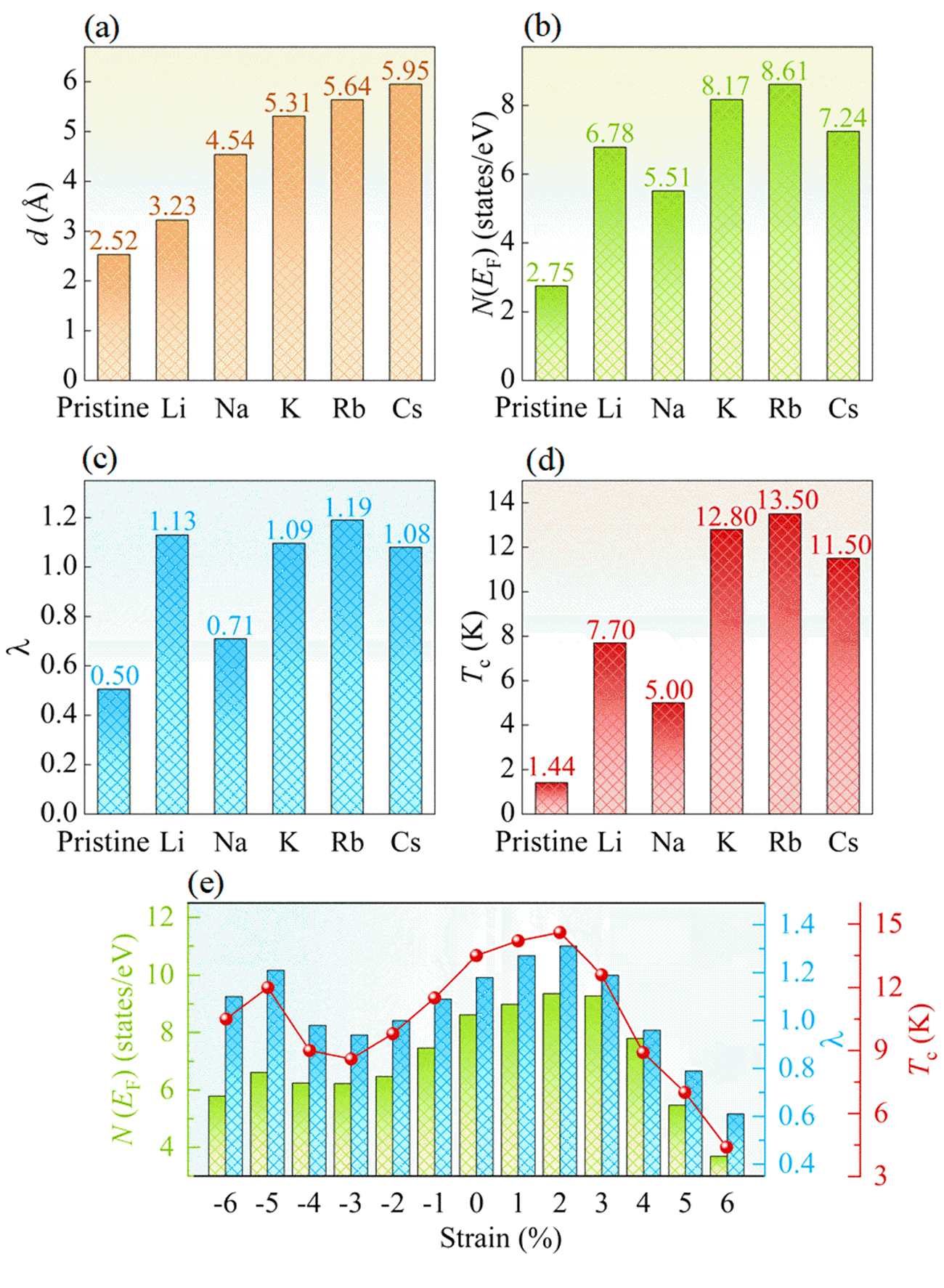}
	\caption{Summarized superconducting parameters. (a) Interlayer spacing $d$, (b) electronic DOS at the Fermi level ($N(E_{F})$), (c) EPC constant $\lambda$, and (d) superconducting $T_{c}$ for pristine and alkali-metal intercalated Pd$_{2}$Te$_{4}$. (e) Corresponding results for RbPd$_{2}$Te$_{4}$ under various biaxial strains.}
	\label{fig:fig6}
\end{figure}

The remaining group-IA intercalants (K, Rb, and Cs) are expected to preserve the single-gap character found in NaPd$_{2}$Te$_{4}$, since the interlayer spacing $d$ increases approximately linearly from Na to K/Rb/Cs intercalation [Fig. 6(a)], consistent with the monotonic increase in the atomic radius. Although K, Rb, and Cs intercalation significantly increases the interlayer spacing (by $\sim$ 2.8--3.4 $\textmd{\AA}$), our calculations indicate pronounced charge transfer and ionic coupling between the intercalants and the adjacent Te layers [Fig. S12]. Moreover, the total energy increases when the interlayer spacing is tuned away from its equilibrium value (either expansion or contraction) [Fig. S13], implying a sizable energetic cost for further separation and therefore this does not support a tendency toward spontaneous exfoliation or phase segregation. Detailed results and discussion can be found in SM \cite{Supplemental-Material}. 
In all three cases, the conduction band dominated metallicity is maintained, while the orbital-resolved FSs exhibit stronger $p$-$d$ hybridization [Fig. S7] \cite{Supplemental-Material}. K and Rb intercalation shift the $E_{F}$ toward a higher vHs peak, yielding $N(E_{F})$ values of 8.17 and 8.60 states/eV [Fig. 4(l) and Fig. 5(b)], respectively. Although Cs intercalation does not fully align $E_{F}$ with the vHs, $E_{F}$ remains very close to it, giving $N(E_{F})$ = 7.24 states/eV [Fig. 5(l)]. This proximity to the vHs offers a promising route to achieving higher $T_{c}$. As expected, the anisotropic results confirm the single-gap nature of K-, Rb-, and Cs-intercalated Pd$_{2}$Te$_{4}$, as evidenced by a single peak in the SDOS [Figs. 4(p)--4(t), Figs. 5(f)--5(j), and Figs. 5(p)--5(t)]. The 2D momentum-resolved maps of $\lambda_{n\mathbf{k}}$ and $\Delta_{n\mathbf{k}}$ on the FSs further show that, with increasing interlayer spacing, the strong-coupling regions and the superconducting gap become concentrated on the FSs around the $K$ point, where the relevant sheets are dominated by hybridized Te-$p$ and Pd-$d$ orbitals [Fig. S7] \cite{Supplemental-Material}. 
The $s$ orbitals of alkali-metal atoms consistently show negligible contributions. In addition, as the mass of the intercalated atoms increases, the strong-coupling region gradually shifts toward low-frequency acoustic branches, particularly the vibrational modes around the $K$ point [Figs. 4(m), 5(c), and 5(m)], consistent with the momentum-resolved distributions of $\lambda_{n\mathbf{k}}$ and $\Delta_{n\mathbf{k}}$. These modes are dominated by in-plane vibrations of Te atoms [Fig. S9] \cite{Supplemental-Material}. 
By integrating $\alpha^{2}F(\omega)$, the resulting $\lambda$ values for KPd$_{2}$Te$_{4}$, RbPd$_{2}$Te$_{4}$, and CsPd$_{2}$Te$_{4}$ are 1.09, 1.19, and 1.08, with corresponding logarithmic average frequencies $\omega_{\log}$ of 104.6, 102.1, and 101.4 K, respectively.
The superconducting transition temperature $T_{c}$ is defined as the temperature at which the normalized $\Delta_{n\mathbf{k}}$ vanishes. Accordingly, RbPd$_{2}$Te$_{4}$ exhibits the highest $T_{c}$ of $\sim$ 13.5 K, while KPd$_{2}$Te$_{4}$ has a comparable value of $\sim$ 12.8 K, followed by CsPd$_{2}$Te$_{4}$ with $T_{c}$ $\sim$ 11.5 K. 

Summarized in Figs. 6(a)--6(d) are the interlayer spacing and several key superconducting parameters of pristine and alkali-metal intercalated Pd$_{2}$Te$_{4}$, including the electronic DOS at the Fermi level $N(E_{F})$, the EPC constant $\lambda$, and the superconducting transition temperature $T_{c}$. As the interlayer spacing increases approximately linearly, the $T_{c}$ values exhibit a nonmonotonic, two-dome-like evolution and correlate closely with the evolution of $N(E_{F})$ and $\lambda$, indicating that the two-dome-like behavior primarily originates from intercalation-induced changes in the electronic structure. 
RbPd$_{2}$Te$_{4}$ exhibits the highest $T_{c}$ among all investigated structures, exceeding the $T_{c}$ values reported in alkali-metal or organic-cations intercalated MoS$_{2}$ \cite{Gate-Controlled-K-Intercalation-and-Superconductivity-in-Molybdenum-Disulfide}, NiTe$_{2}$ \cite{Emergent-superconductivity-in-two-dimensional-NiTe2-crystals}, PtTe$_{2}$ \cite{Enhanced-superconductivity-in-bilayer-PtTe2-by-alkali-metal-intercalations}, and SnSe$_{2}$ \cite{Superconductivity-in-Li-intercalated-1T-SnSe2-driven-by-electric-field-gating,Induced-anisotropic-superconductivity-in-ionic-liquid-cation-intercalated-1T-SnSe2,Multiple-Intercalation-Stages-and-Universal-Tc-Enhancement-through-Polar-Organic-Species-in-Electron-Doped-1T-SnSe2}.
Especially, compared to PdTe$_{2}$ across different thicknesses ($T_{c}$ $\sim$ 0$-$1.8 K) \cite{Two-dimensional-superconductivity-and-topological-states-in-PdTe2-thin-films,Ionic-liquid-gating-induced-self-intercalation-of-transition-metal-chalcogenides,Electron-phonon-coupling-in-superconducting-1T-PdTe2,The-occurrence-of-superconductivity-in-sulfides-selenides-tellurides-of-Pt-group-metals,Constitution-and-magnetic-and-electrical-properties-of-palladium-tellurides-PdTe-PdTe2,Superconductivity-in-Cu-Intercalated-CdI2-Type-PdTe2,Protonation-enhanced-superconductivity-in-PdTe2}, alkali-metal intercalation enhances $T_{c}$ by up to an order of magnitude. This improvement highlights that intercalation offers an additional and highly effective tuning parameter beyond thickness control.
Moreover, the experimental realization of these 2D superconductors appears feasible. Given the mature synthesis techniques for TMDCs, PdTe$_{2}$ thin films with different layer numbers have already been successfully fabricated \cite{Two-dimensional-superconductivity-and-topological-states-in-PdTe2-thin-films,Type-II-Ising-Superconductivity-and-Anomalous-Metallic-State-in-Macro-Size-Ambient-Stable-Ultrathin-Crystalline-Films}, and alkali-metal intercalation into bilayers can be achieved via ionic-liquid gating \cite{Ionic-liquid-gating-induced-self-intercalation-of-transition-metal-chalcogenides,Gate-tunable-phase-transitions-in-thin-flakes-of-1T-TaS2,Gate-induced-superconductivity-in-a-monolayer-topological-insulator}. Therefore, alkali-metal-intercalated Pd$_{2}$Te$_{4}$ thin films with relatively higher $T_{c}$ are also expected to be experimentally synthesized.

Considering that experimental intercalation may slightly deviate from the ideal stoichiometry $n$ = 0.5, we test the robustness of the predicted two-gap states in Li-intercalated bilayer using a 3 $\times$ 3 $\times$ 1 supercell based on Li$_{9}$(Pd$_{2}$Te$_{4}$)$_{9}$. Li vacancy and Li excess are modeled by removing or adding one Li atom, yielding Li$_{8}$(Pd$_{2}$Te$_{4}$)$_{9}$ and Li$_{10}$(Pd$_{2}$Te$_{4}$)$_{9}$, respectively. Although fully converged anisotropic ME calculations for these supercells are computationally prohibitive, the two-gap to single-gap transition is governed by whether the $p_{z}$-like band crosses $E_{F}$, i.e., whether the corresponding $p_{z}$-like Fermi pocket exists. Accordingly, we calculate the electronic band structures of Li$_{8}$(Pd$_{2}$Te$_{4}$)$_{9}$ and Li$_{10}$(Pd$_{2}$Te$_{4}$)$_{9}$ and unfold them to the first BZ for direct comparison with the $n$ = 0.5 primitive cell (LiPd$_{2}$Te$_{4}$). For the Li-vacancy case (hole-doped), the $p_{z}$-like band still crosses $E_{F}$ [Fig. S14(a)] \cite{Supplemental-Material}, suggesting that the $p_{z}$-like pocket persists and the two-gap character should be retained. Moreover, a higher vacancy concentration would further shift $E_{F}$ downward, tending to maintain the $p_{z}$-like band crossing and thus supporting the robustness of the two-gap character against Li vacancies. In contrast, for the Li-excess (electron-doped) case Li$_{10}$(Pd$_{2}$Te$_{4}$)$_{9}$, the extra Li shifts $E_{F}$ upward such that it no longer intersects the $p_{z}$-like band [Fig. S14(b)] \cite{Supplemental-Material}, and the system is therefore expected to transition to a single-gap state. It should be emphasized that the Li-excess level in Li$_{10}$(Pd$_{2}$Te$_{4}$)$_{9}$ corresponds to a relatively large deviation ($\sim$ 11.1\%) within the present 3 $\times$ 3 $\times$ 1 supercell model. Given the continuous evolution of the electronic structure with doping, these results suggest a critical excess level beyond which the $p_{z}$-like pocket vanishes. Therefore, the two-gap feature may still persist for slight Li-excess levels, while larger excess drives the system to undergo the transition. Overall, the transition is governed by a well-defined electronic criterion, namely, the presence or absence of the $p_{z}$-like Fermi-surface pocket, and the two-gap superconducting characteristic in LiPd$_{2}$Te$_{4}$ is relatively robust against slight stoichiometric deviations. Additionally, for heavier alkali-metal (Na, K, Rb, and Cs) intercalated Pd$_{2}$Te$_{4}$, the $p_{z}$-like valence band lies far below the Fermi level. Therefore, slight stoichiometric deviations of alkali metals may insufficient to drive this band back across $E_{F}$. As a result, the single-gap character in these systems is expected to be robust. Detailed results can be found in SM \cite{Supplemental-Material}.

\begin{figure*}
	\centering
	\includegraphics[width=0.85\linewidth]{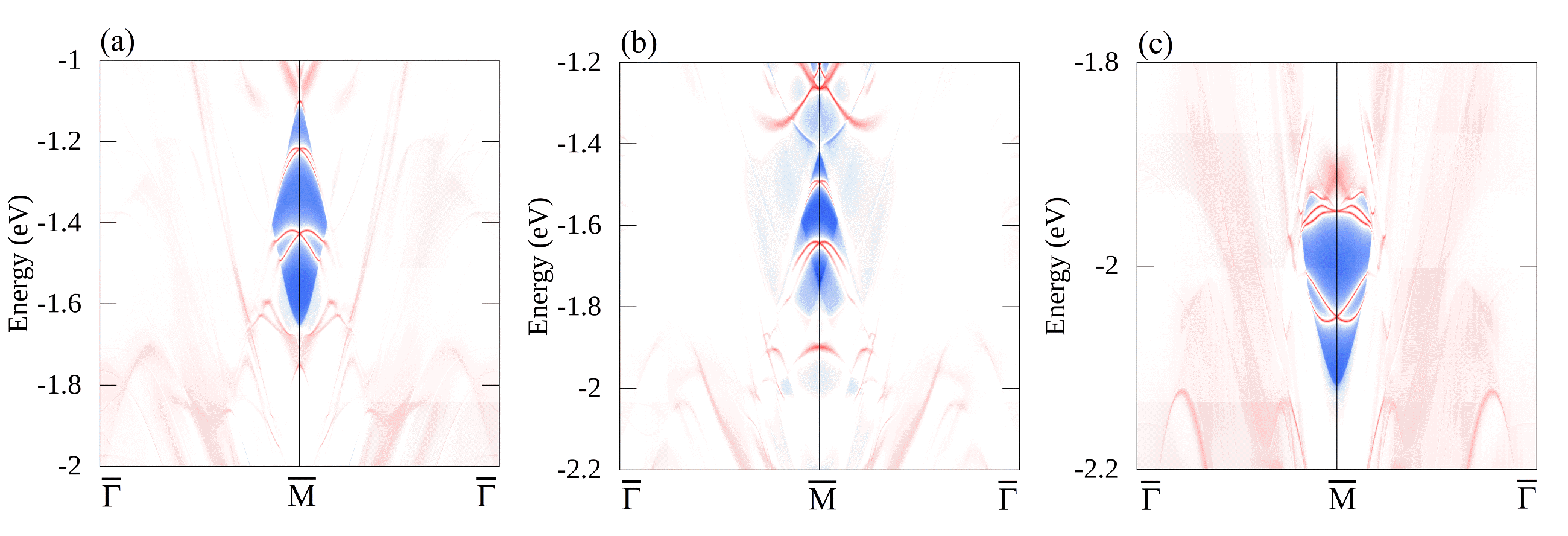}
	\caption{Edge states along $\overline{\Gamma}$-$\overline{M}$-$\overline{\Gamma}$ direction for (a) Pd$_{2}$Te$_{4}$,  (b) LiPd$_{2}$Te$_{4}$, and (c) NaPd$_{2}$Te$_{4}$, weighted by the local DOS. The $E_{F}$ is set to zero.}
	\label{fig:fig7}
\end{figure*}

\subsection{Strain induced two-dome-like superconductivity in RbPd$_{2}$Te$_{4}$}
Given possible in-plane strains in ultrathin materials, which can be induced by substrate, and the potential discrepancies in lattice constants between calculations and experimental measurements, it is necessary to investigate how the EPC and superconductivity in this system evolve under in-plane strain. Here, we focus on RbPd$_{2}$Te$_{4}$, which exhibits the highest superconducting $T_{c}$, and apply biaxial strain by adjusting the in-plane lattice constant and then optimizing the internal coordinates to assess the strain effect. The biaxial strain is defined as $\epsilon$ = ($a$$-$$a_{0}$)/$a_{0}$, where $a_{0}$ is the optimized lattice constant of unstrained RbPd$_{2}$Te$_{4}$ and $a$ is the corresponding lattice constant under strain. 
Presented in Fig. 6(e) is the evolution of $N(E_{F})$, $\lambda$, and superconducting $T_{c}$ as a function of biaxial strain, as obtained by solving the fully anisotropic ME equations. Specifically, as the biaxial tensile strain increases, both $N(E_{F})$ and $\lambda$ exhibit a dome-like variation, first rising and then falling, whereas the opposite trend is observed under biaxial compressive strain. 
Overall, the $T_{c}$ of RbPd$_{2}$Te$_{4}$ exhibits a notable two-dome-like dependence on biaxial strain, reaching a maximum of $\sim$ 14.5 K at 2\% tensile strain, with the corresponding $N(E_{F})$ and $\lambda$ of 9.36 states/eV and 1.31, respectively. Detailed orbital-resolved electronic and phonon properties, as well as anisotropic superconducting results for RbPd$_{2}$Te$_{4}$ under 2\% tensile strain are presented in Fig. S18 \cite{Supplemental-Material}. Further increasing the biaxial strain beyond 6\% is expected to weaken the EPC and may induce dynamical instability in the RbPd$_{2}$Te$_{4}$ crystal \cite{Supplemental-Material}. 

The two-dome-like behavior can be well rationalized by the strain-induced evolution of the electronic structure. Specifically, under small tensile strain, the two conduction bands crossing  $E_{F}$ become less dispersive, which enhances the vHs in electronic DOS and leads to an increase in $T_{c}$. 
Although further tensile strain beyond 2\% [Fig. S16] \cite{Supplemental-Material} makes the conduction bands increasingly localized, the bands along the $M$--$K$--$\Gamma$ direction are shifted away from $E_{F}$, resulting in a reduced DOS and consequently weakening the EPC and lowering $T_{c}$, thereby giving rise to the dome-like dependence under biaxial tensile strain in RbPd$_{2}$Te$_{4}$.
However, under biaxial compressive strain the situation is different. The conduction bands become more dispersive as the compressive strain increases, which reduces $N(E_{F})$ and is accompanied by a decreased separation between the valence and conduction bands. 
This explains the decrease in $T_{c}$ in the initial compressive-strain regime (1\%--3\%). When the applied compressive strain exceeds 3\%, the band-gap narrows and the $\Gamma$-centered $p_{xy}$-like valence band becomes metallic [Fig. S16] \cite{Supplemental-Material}, thereby increasing the electronic DOS. At 5\% compressive strain, $E_{F}$ is tuned toward the vHs, yielding $N(E_{F})$ = 6.61 (states/eV) and the highest $T_{c}$, with $\lambda$ = 1.21 and $T_{c}$ = 12.0 K, respectively.
Further increasing the compressive strain will shift $E_{F}$ away from the vHs again, weakening superconductivity and ultimately giving rise to the dome-like $T_{c}$ dependence under biaxial compressive strain in RbPd$_{2}$Te$_{4}$ [Fig. S17] \cite{Supplemental-Material}. This also suggests that the substrate with slightly larger lattice constant may help to promote the $T_{c}$. 
Additionally, the strain range considered here (up to ±6\%) is relatively moderate, as even larger biaxial strains have been explored in 2D layered systems. From a theoretical perspective, numerous studies have applied biaxial strains exceeding 6\%\cite{First-Principles-Calculations-on-the-Effect-of-Doping-and-Biaxial-Tensile-Strain-on-Electron-Phonon-Coupling-in-Graphene,Electron-phonon-driven-charge-density-wave-and-superconductivity-in-a-1TTaSi2N4-monolayer,Superconductivity-and-strain-enhanced-phase-stability-of-Janus-tungsten-chalcogenide-hydride-monolayers}. Experimentally, tensile strain up to about 25\% has been elastically applied to graphene without breaking \cite{Measurement-of-the-Elastic-Properties-and-Intrinsic-Strength-of-Monolayer-Graphene}. Moreover, the extraordinary flexibility and mechanical properties of 2D TMDCs enable them to tolerate substantial structural curvature and accommodate large deformations in both the in-plane and out-of-plane directions. Specifically, 2D TMDCs can withstand up to 10\% strain, which is an order of magnitude higher than their bulk counterparts \cite{Brittle-Fracture-of-2D-MoSe2,Mechanical-Anisotropy-in-Two-Dimensional-Selenium-Atomic-Layers,Mechanical-Properties-of-Atomically-Thin-Tungsten-Dichalcogenides-WS2-WSe2-and-WTe2,Local-Strain-Engineering-of-Two-Dimensional-Transition-Metal-Dichalcogenides-Towards-Quantum-Emitters}. Therefore, adopting a ±6\% strain range in this work provides a reasonable window to establish strain-dependent trends while remaining within a commonly used scope.

\subsection{Nontrivial electronic band topology}	
It is worth noting that rich topological phases and Dirac states have been reported in bulk PdTe$_{2}$ \cite{Experimental-Realization-of-Type-II-Dirac-Fermions-in-a-PdTe2-Superconductor,Fermiology-and-Superconductivity-of-Topological-Surface-States-in-PdTe2,Identification-of-Topological-Surface-State-in-PdTe2-Superconductor-by-Angle-Resolved-Photoemission-Spectroscopy}. Previous studies have further shown that Dirac cones and nontrivial band topology are retained in ML and few-layer PdTe$_{2}$ \cite{Two-dimensional-superconductivity-and-topological-states-in-PdTe2-thin-films,Observation-of-Gapped-Topological-Surface-States-and-Isolated-Surface-Resonances-in-PdTe2-Ultrathin-Films}. 
This motivates us to examine whether the nontrivial band topology can be preserved in alkali-metal intercalated bilayer PdTe$_{2}$, which may provide a platform for exploring systems that combine a higher superconducting $T_{c}$ with topological characteristics. 

Therefore, by analyzing the band structures of alkali-metal intercalated Pd$_{2}$Te$_{4}$, we find that including spin-orbit coupling eliminates the degeneracy of the Dirac cone and opens finite energy gaps, as shown in Fig. S20 \cite{Supplemental-Material}. 
We then evaluate the ${Z}_{2}$ invariant to characterize the electronic topology. It is 1 for pristine and Li/Na-intercalated Pd$_{2}$Te$_{4}$, but becomes 0 for the remaining alkali-metal intercalations (K, Rb, and Cs).
This indicates that the band topology undergoes a transition from nontrivial to trivial as the intercalation evolves from Li/Na to heavier alkali metals. Combined with the superconductivity results, these trends suggest that a tunable regime in which superconductivity coexists with nontrivial band topology is most likely realized under moderate doping and/or interlayer expansion (i.e., Li/Na-like conditions). 
By contrast, heavier intercalation, while typically enhancing $T_{c}$, tends to push the system into a topologically trivial state, highlighting an intrinsic trade-off that could be optimized by carefully controlling the carrier doping concentration and interlayer interaction.
Furthermore, Figs. 7(a)--7(c) clearly display Rashba-like topological edge states for all three structures, thereby identifying them as ${Z}_{2}$ topological metals. These edge states can be experimentally detected by angle-resolved photoemission spectroscopy. Further low-temperature experiments are anticipated to confirm the existence of these 2D nontrivial topological edge states in alkali-metal intercalated Pd$_{2}$Te$_{4}$.

\section{Conclusion}
In conclusion, we employed first-principles calculations in conjunction with the Wannier interpolation technique to investigate the electronic structure, electron-phonon coupling, and anisotropic superconducting properties of Pd$_{2}$Te$_{4}$. The results indicate that the weak superconductivity of Pd$_{2}$Te$_{4}$ can be significantly enhanced by group-IA alkali-metal intercalation. Notably, the $T_{c}$ of intercalated Pd$_{2}$Te$_{4}$ exhibits a nonmonotonic dependence on the intercalant, with Rb intercalation yielding the highest $T_{c}$ of $\sim$ 13.5 K. A further enhancement to $\sim$ 14.5 K can be achieved under biaxial tensile strain in RbPd$_{2}$Te$_{4}$, where the strain-dependent $T_{c}$ exhibits an intriguing two-dome-like behavior. More importantly, this work reveals a systematic correlation between the superconducting gap and interlayer coupling in alkali-metal intercalated Pd$_{2}$Te$_{4}$. Li intercalation induces a distinct two-gap superconducting state, whereas intercalants with larger atomic radii (Na, K, Rb, and Cs) drive the system into a single-gap character. This transition can be attributed to the modulation of interlayer coupling strength via intercalation-induced interlayer expansion in Pd$_{2}$Te$_{4}$. Moreover, pristine and Li/Na-intercalated bilayers exhibit nontrivial band topology, suggesting that layered PdTe$_{2}$ provides a promising platform for realizing the coexistence of superconductivity and nontrivial topology. These results provide detailed anisotropic insights into EPC and offer viable pathways for enhancing $T_{c}$ and achieving diverse properties in layered PdTe$_{2}$ systems.
		
\vspace{0.5cm}
\begin{acknowledgements}
This work is supported by the National Natural Science Foundation of China (Grant No. 12074213, No. 11574108, No. U2330104, and No. 12574028), the National Key R$\&$D Program of China (Grant No. 2022YFA1403103), the Major Basic Program of Natural Science Foundation of Shandong Province (Grant No. ZR2021ZD01), and the Project of Introduction and Cultivation for Young Innovative Talents in Colleges and Universities of Shandong Province.
	
\end{acknowledgements}
%


	\end{spacing}
\end{document}